\documentclass[a4paper,11pt]{article}
\usepackage{pos}
\usepackage{xcolor}
\usepackage{subfig}

\newcommand\psibar{\overline{\psi}}
\newcommand\avg[1]{\left\langle {#1} \right\rangle}
\DeclareMathOperator{\Tr}{Tr}

\usepackage{pgfplotstable}
\usepackage{booktabs}
\usepackage{multirow}

\definecolor{darkcandyapplered}{rgb}{0.64, 0.0, 0.0}
\definecolor{goethe}{rgb}{0,0.37,0.66}

\pgfplotsset{compat=1.17}
\pgfplotstableread[col sep=comma]{
nf,name,nt,mean,errorp,errorm,reference
,std,4,285,7.5,7.5,\cite{karschChiralCriticalPoint2001a}
,std,4,270,17,17,\cite{christLocating3flavorCritical2003b}
3,std,6,147,4,4,\cite{deforcrandQCDChiralCritical2008}
,p4,4,67,9,9,\cite{karschWhereChiralCritical2004a}
,HISQ,6,50,0,10,\cite{bazavovChiralPhaseStructure2017a}
2,std-$\mu_i$,4,60,12,12,\cite{bonatiChiralPhaseTransition2014a}
}\dataStaggered

\pgfplotstableread[col sep=comma]{
nf,name,nt,mean,errorp,errorm,reference
3,clover,6-8,304,14,14,\cite{jinCriticalEndpointFinite2015}
,clover,8-10,170,0,34,\cite{jinCriticalPointPhase2017}
,clover,10-12,110,0,22,\cite{kuramashiNaturePhaseTransition2020}
2,std,4,560,3,3,\cite{philipsenQCDChiralPhase2016a}
,tm,12,300,0,300,\cite{burgerThermalQCDTransition2013b}
,clover,16,200,0,200,\cite{brandtStrengthUAAnomaly2016}
}\dataWilson

\newcommand{\ZTwoUniversality}{Z_2}
\tikzset {
    halo/.style= { 
        preaction= {
            draw,
            white,
            line width=7,
            -
        }}}

\newcommand{\errplotWilson}{%
  \begin{tikzpicture}[trim axis left,trim axis right]
    \begin{axis}[y=-\baselineskip,
        scale only axis,
        width             = 6.5cm,
        enlarge y limits  = {abs=0.5},
        axis y line*      = middle,
        y axis line style = dashed,
        ytick             = \empty,
        axis x line*      = bottom,
        xtick             = {0, 200, 400, 600}
      ]
      \addplot+[only marks, color=goethe][error bars/.cd,x dir=both, x explicit]
        table [x=mean,y expr=\coordindex,x error plus=errorp,x error minus=errorm]{\dataWilson};
    \end{axis}
  \end{tikzpicture}%
}
\newcommand{\errplotStaggered}{%
  \begin{tikzpicture}[trim axis left,trim axis right]
    \begin{axis}[y=-\baselineskip,
        scale only axis,
        width             = 6.5cm,
        enlarge y limits  = {abs=0.5},
        axis y line*      = middle,
        y axis line style = dashed,
        ytick             = \empty,
        axis x line*      = bottom,
        xtick             = {0, 100, 200, 300}
      ]
      \addplot+[only marks, color=goethe][error bars/.cd,x dir=both, x explicit]
        table [x=mean,y expr=\coordindex,x error plus=errorp,x error minus=errorm]{\dataStaggered};
    \end{axis}
  \end{tikzpicture}%
}

\title{QCD thermodynamics with stabilized Wilson fermions}

\author*[a]{Rocco Francesco Basta}
\author[b]{Bastian B. Brandt}
\author[a]{Francesca Cuteri}
\author[b]{Gergely Endrődi}
\author[c]{Anthony Francis}

\affiliation[a]{Institut für Theoretische Physik, Goethe-Universität Frankfurt,\\
  Max-von-Laue-Straße 1, 60438 Frankfurt am Main, Germany}

\affiliation[b]{Institute for Theoretical Physics, University of Bielefeld, D-33615 Bielefeld, Germany}

\affiliation[c]{Institute of Physics, National Yang Ming Chiao Tung University, 30010 Hsinchu, Taiwan}

\emailAdd{basta@itp.uni-frankfurt.de}

\abstract{Stabilized Wilson fermions are a reformulation of Wilson clover fermions that incorporates several numerical stabilizing techniques, but also a local change of the fermion action -- the original clover term being replaced with an exponentiated version of it. We intend to apply the stabilized Wilson fermions toolbox to the thermodynamics of QCD, starting on the $N_f=3$ symmetric line on the Columbia plot, and to compare the results with those obtained with other fermion discretizations.}

\FullConference{%
The 39th International Symposium on Lattice Field Theory,\\
8th-13th August, 2022,\\
Rheinische Friedrich-Wilhelms-Universität Bonn, Bonn, Germany
}

\begin{document}
\maketitle

\section{Introduction}

\begin{figure}
    \centering
    \subfloat[]{\includegraphics[width=0.4\textwidth]{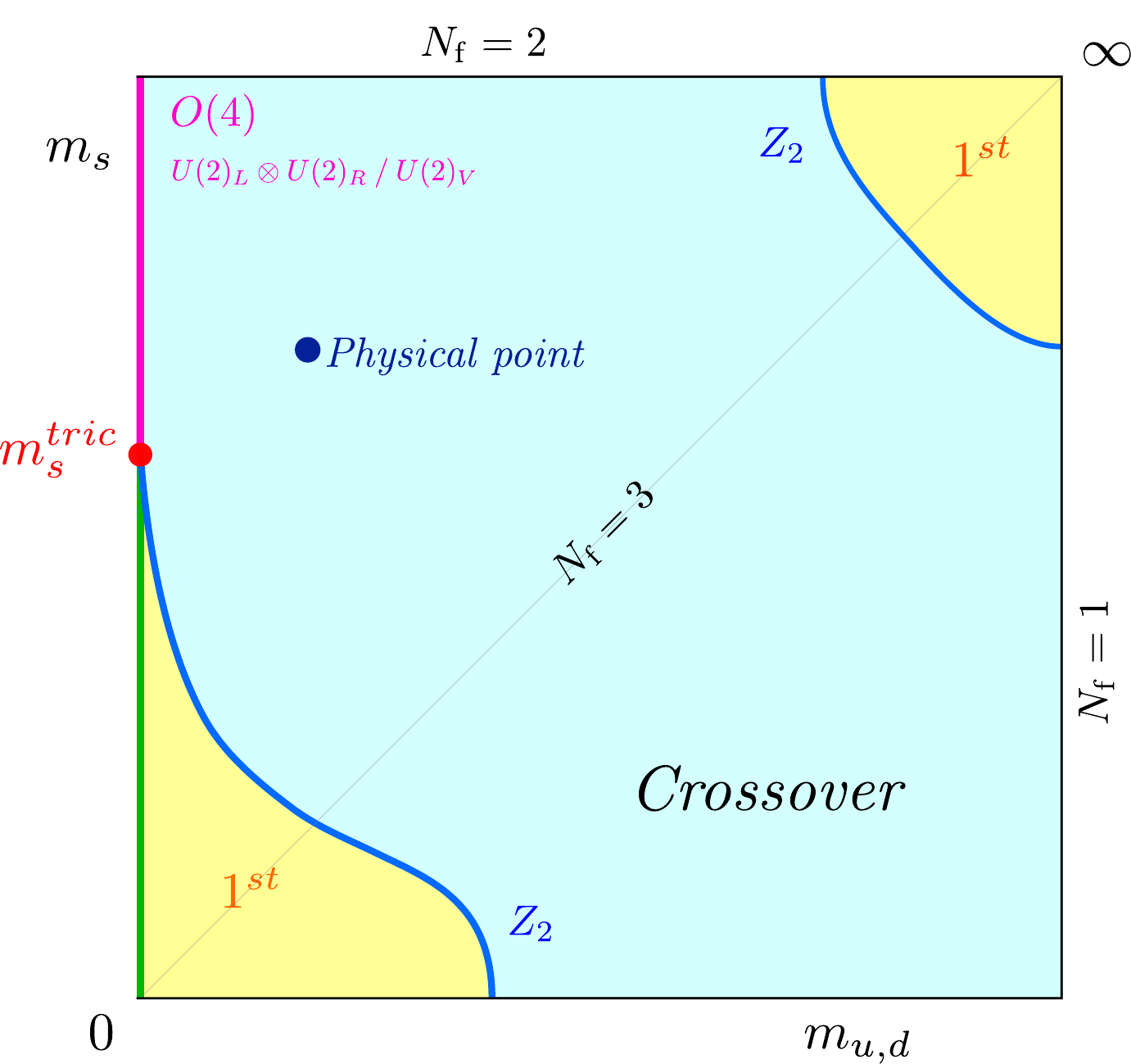}}
    \subfloat[]{\includegraphics[width=0.4\textwidth]{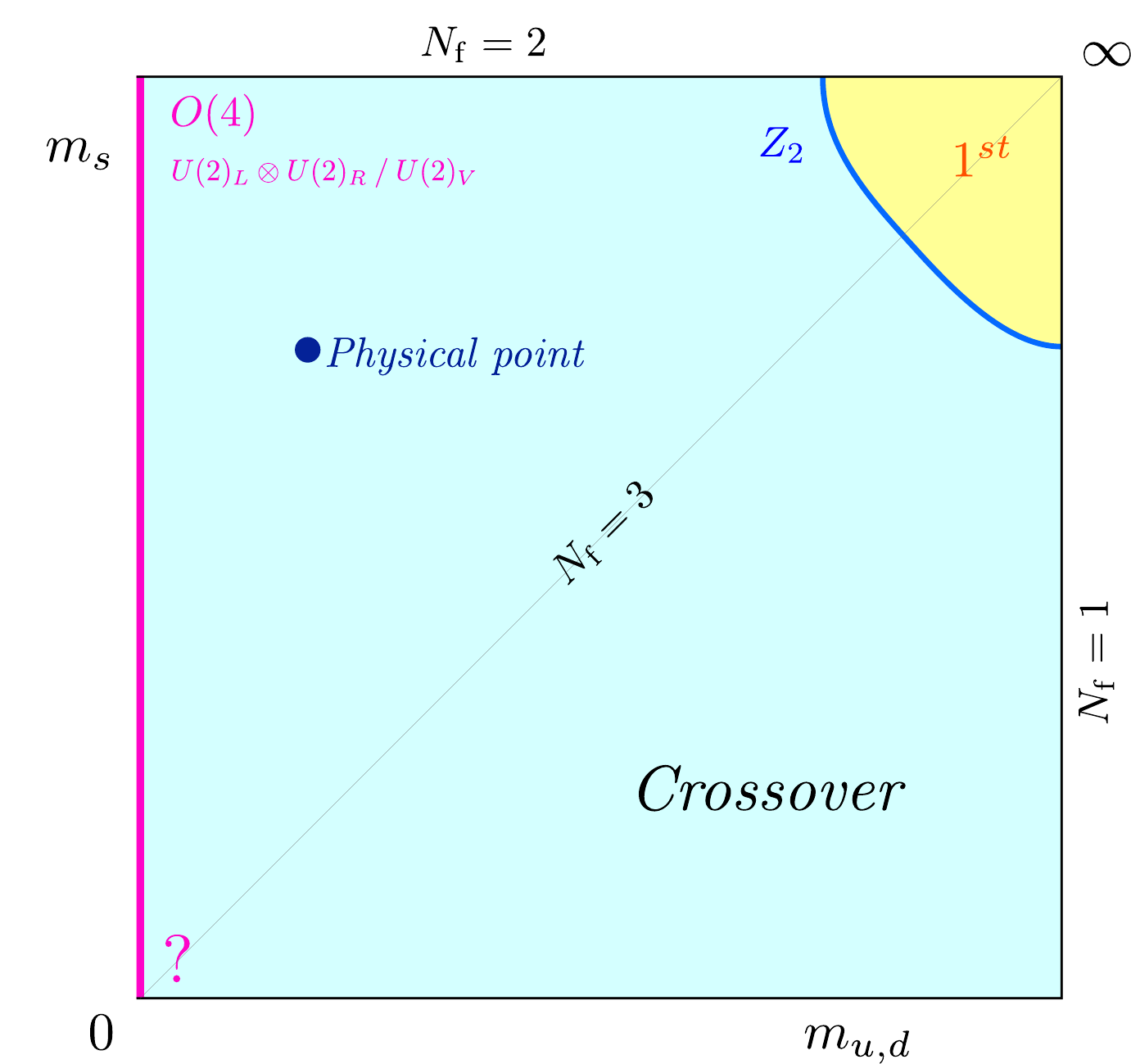}}
    \caption{Two possible scenarios for the light quark mass region of the Columbia plot in the continuum limit: either (a) the first-order region remains finite, ending with a tricritical point, or (b) the first-order region vanishes in the continuum limit. Figures taken from Ref.~\cite{cuteriOrderQCDChiral2021}.}
    \label{fig:columbia_plot}
\end{figure}

The order of the thermal phase transition (Fig.~\ref{fig:columbia_plot}) in QCD depends on the number of quark flavours and on their masses. While the transition is a smooth crossover at the physical point, a first order transition region \cite{cuteriDeconfinementCriticalPoint2021} was found when the light and strange quark masses are sufficiently heavy. The crossover and first order regions are separated by a second order critical line in the $Z_2$ universality class. As for what precisely happens near the chiral limit, the picture is less clear.
When approaching the chiral limit with $N_f = 2$, evidence for a second order $O(4)$ transition line ending with a tricritical point has been recently found \cite{cuteriQCDChiralPhase2018, cuteriOrderQCDChiral2021}. Performing a similar study on the $N_f = 3$ flavour symmetric line, a first-order region \cite{gavaiMetastabilitiesThreeflavorQCD1987}, again separated by a second-order $Z_2$ line \cite{karschWhereChiralCritical2004a}, appears. A way to quantify the extent of this region is to measure the pion mass $m_\pi^{Z_2}$ corresponding to the intersection between the second-order $Z_2$ critical line and the $N_f=3$ diagonal. 

The precise location of this $Z_2$ endpoint in the Columbia plot is still unclear. In the literature (Tab.~\ref{tab:z2}), there are many results obtained with both staggered and Wilson fermions, and different types of related improved actions, but they cover a wide range of possible masses. Cutoff effects are found to play a significant role in the light mass sector of the plot, and it is even possible that the first order region could vanish in the continuum limit. This would have profound implications for our understanding of the phase diagram at finite baryon chemical potential \cite{rajagopalCondensedMatterPhysics2001}, which so far cannot be simulated on the lattice due to the sign problem.

\begin{figure}[h]
    \centering
    \includegraphics[width=0.5\textwidth]{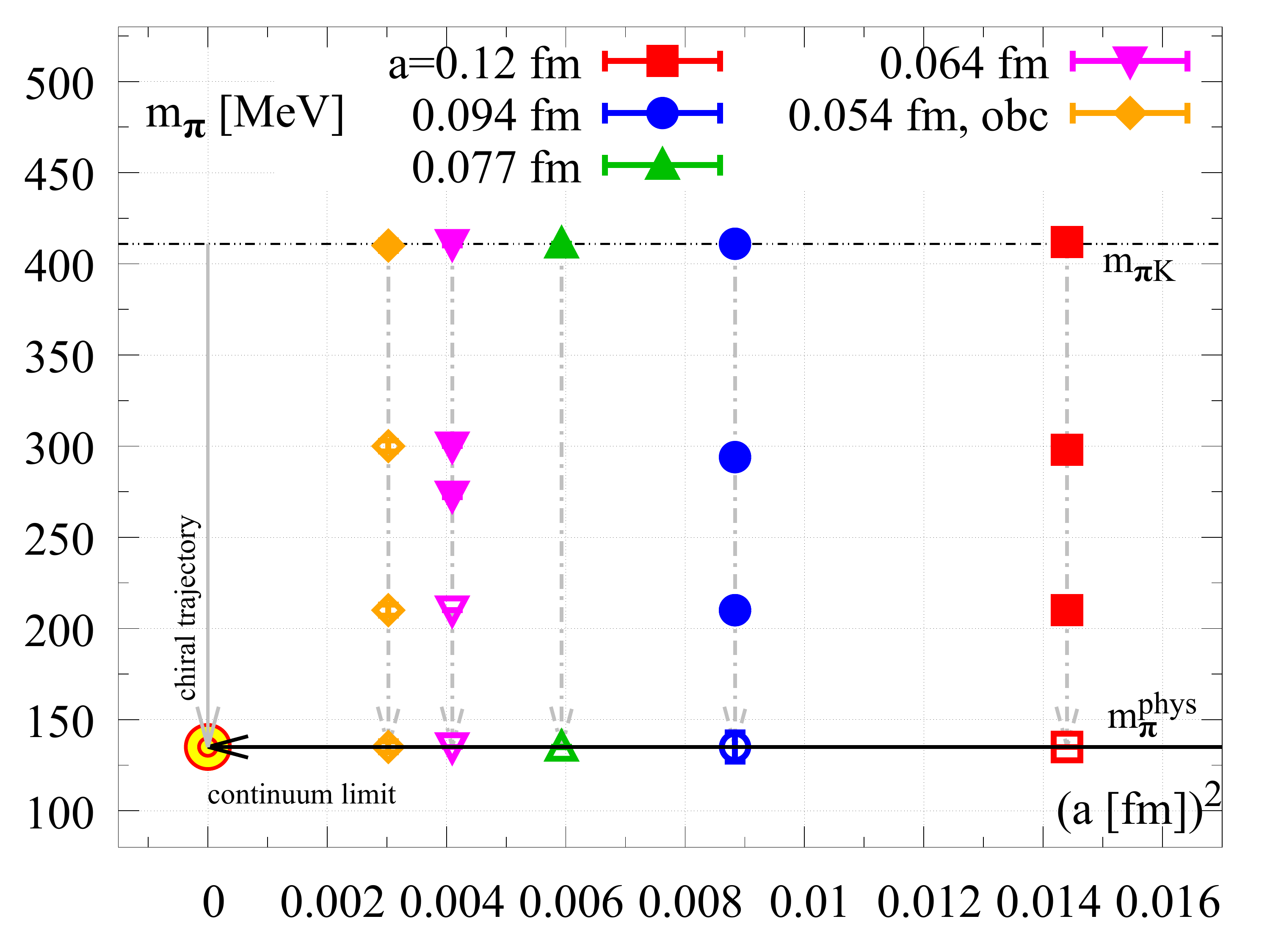}
    \caption{Ensembles being tuned and generated by the Open Lattice initiative. Figure taken from Ref.~\cite{cuteriPropertiesEnsemblesHadron2022}.}
    \label{fig:oli_ensembles}
\end{figure}

Simulating QCD at low quark masses is a challenging task and various techniques have been developed to deal with the increasing condition number of the Dirac operator. Improved actions are necessary to reduce cutoff effects, yet Wilson Clover fermions are known to cause instabilities at light pion masses, because the lowest eigenvalue of the Dirac operator can become arbitrarily small.
Stabilized Wilson fermions \cite{francisMasterfieldSimulationsImproved2020,cuteriPropertiesEnsemblesHadron2022} have recently been developed as an alternative to the standard $O(a)$-improved Wilson Clover action, with the attempt to mitigate some of its drawbacks. Multiple zero-temperature ensembles are being generated by the Open Lattice initiative \cite{cuteriPropertiesEnsemblesHadron2022} (Fig.~\ref{fig:oli_ensembles}) with an open science philosophy underlying its efforts. Our intent is to apply these tools to the study of QCD thermodynamics, starting with the ensembles currently available on the flavour symmetric $N_f=3$ line, and investigate their impact on the location of the $Z_2$ endpoint.

\begin{table}
    \centering
    \subfloat[Results obtained with staggered fermions]{
\pgfplotstablegetrowsof{\dataStaggered}
\let\numberofrows=\pgfplotsretval

\pgfplotstabletypeset[columns={nf,name,nt,errorp,reference},
  every head row/.style = {before row=\toprule, after row=\midrule},
  every last row/.style = {after row=[3ex]\bottomrule},
  columns/nf/.style = {string type, column name={$N_{\mathrm{f}}$}},
  columns/name/.style = {string type, column name=Action},
  columns/nt/.style = {string type, column name={$N_\tau$}},
  columns/reference/.style = {string type, column name={Ref.}},
  columns/errorp/.style = {
    column name = {$m_\pi^{\ZTwoUniversality}$ [MeV]},
    assign cell content/.code = {
    \ifnum\pgfplotstablerow=0
    \pgfkeyssetvalue{/pgfplots/table/@cell content}
    {\multirow{\numberofrows}{6.5cm}{\errplotStaggered}}%
    \else
    \pgfkeyssetvalue{/pgfplots/table/@cell content}{}%
    \fi
    }
  }
]{\dataStaggered}
    }\\
    \subfloat[Results obtained with Wilson fermions]{
\pgfplotstablegetrowsof{\dataWilson}
\let\numberofrows=\pgfplotsretval

\pgfplotstabletypeset[columns={nf,name,nt,errorp,reference},
  every head row/.style = {before row=\toprule, after row=\midrule},
  every last row/.style = {after row=[3ex]\bottomrule},
  columns/nf/.style = {string type, column name={$N_{\mathrm{f}}$}},
  columns/name/.style = {string type, column name=Action},
  columns/nt/.style = {string type, column name={$N_\tau$}},
  columns/reference/.style = {string type, column name={Ref.}},
  columns/errorp/.style = {
    column name = {$m_\pi^{\ZTwoUniversality}$ [MeV]},
    assign cell content/.code = {
    \ifnum\pgfplotstablerow=0
    \pgfkeyssetvalue{/pgfplots/table/@cell content}
    {\multirow{\numberofrows}{6.5cm}{\errplotWilson}}%
    \else
    \pgfkeyssetvalue{/pgfplots/table/@cell content}{}%
    \fi
    }
  }
]{\dataWilson}
    }
    \caption{Results in the literature for the location of the $Z_2$ endpoints in the light-quark region for $N_f=2$ and $3$, with various fermionic actions. Note that $N_f=3$ results cover a range that goes from $50$ MeV to $304$ MeV.}
    \label{tab:z2}
\end{table}

\section{Stabilized Wilson fermions}

First developed in order to cure a pathology of Wilson-Clover fermions in large volumes \cite{francisMasterfieldSimulationsImproved2020}, stabilized Wilson fermions have been shown \cite{cuteriPropertiesEnsemblesHadron2022} to also benefit ordinary simulations by reducing cutoff effects, potentially enabling the use of coarser lattices, and by reducing the probability of encountering exceptional configurations. They consist in

\begin{itemize}
    \item Stochastic Molecular Dynamics (SMD) update algorithm, which has been shown to be less affected by instabilities in the molecular dynamics (MD) evolution than standard HMC;
    \item An exponentiated version of the Clover improvement term, which protects its contribution to the Dirac operator from arbitrarily small eigenvalues and is guaranteed to be safely invertible;
    \item Uniform norm solver stopping criterion, to protect from precision loss from local effects when large lattices are used;
    \item Quadruple precision arithmetic, in order to further reduce precision loss effects.
\end{itemize}
These features are implemented in the open source software \texttt{openQCD-2.0} and higher \cite{luscherOpenQCD}.

\section{Methodology}

When approaching the chiral limit, the light chiral condensate, defined as 
\begin{equation}
    \langle \psibar \psi \rangle = \frac{T}{V} \frac{\partial \log Z}{\partial m_{ud}} = \frac{T}{V} \avg{\Tr (D^{-1})}
\end{equation}
can be regarded as an approximate order parameter for the thermal phase transition. The statistical properties of the sampled $\avg{\psibar \psi}$ distribution can be studied to obtain information on the phase transition. This means studying the (disconnected) chiral susceptibility
    \begin{equation}
        \chi_{\psibar\psi} = \frac{T}{V} \left[ \avg{\Tr (D^{-1})^2} - \avg{\Tr (D^{-1})}^2 \right]
    \end{equation}
and the $n$-th standardized moments of the chiral condensate
        \begin{equation}
            B_n (\beta, \kappa, N_\sigma) = \frac{\avg{(\psibar \psi - \avg{\psibar \psi})^n}}{\avg{(\psibar \psi - \avg{\psibar \psi})^2}^{n/2}}.
        \end{equation}
where $V \equiv (N_\sigma a)^3$ is the spatial volume.

\begin{figure}
    \centering
    \subfloat[Skewness\label{fig:moments_skewness}]{\includegraphics[width=0.5\textwidth]{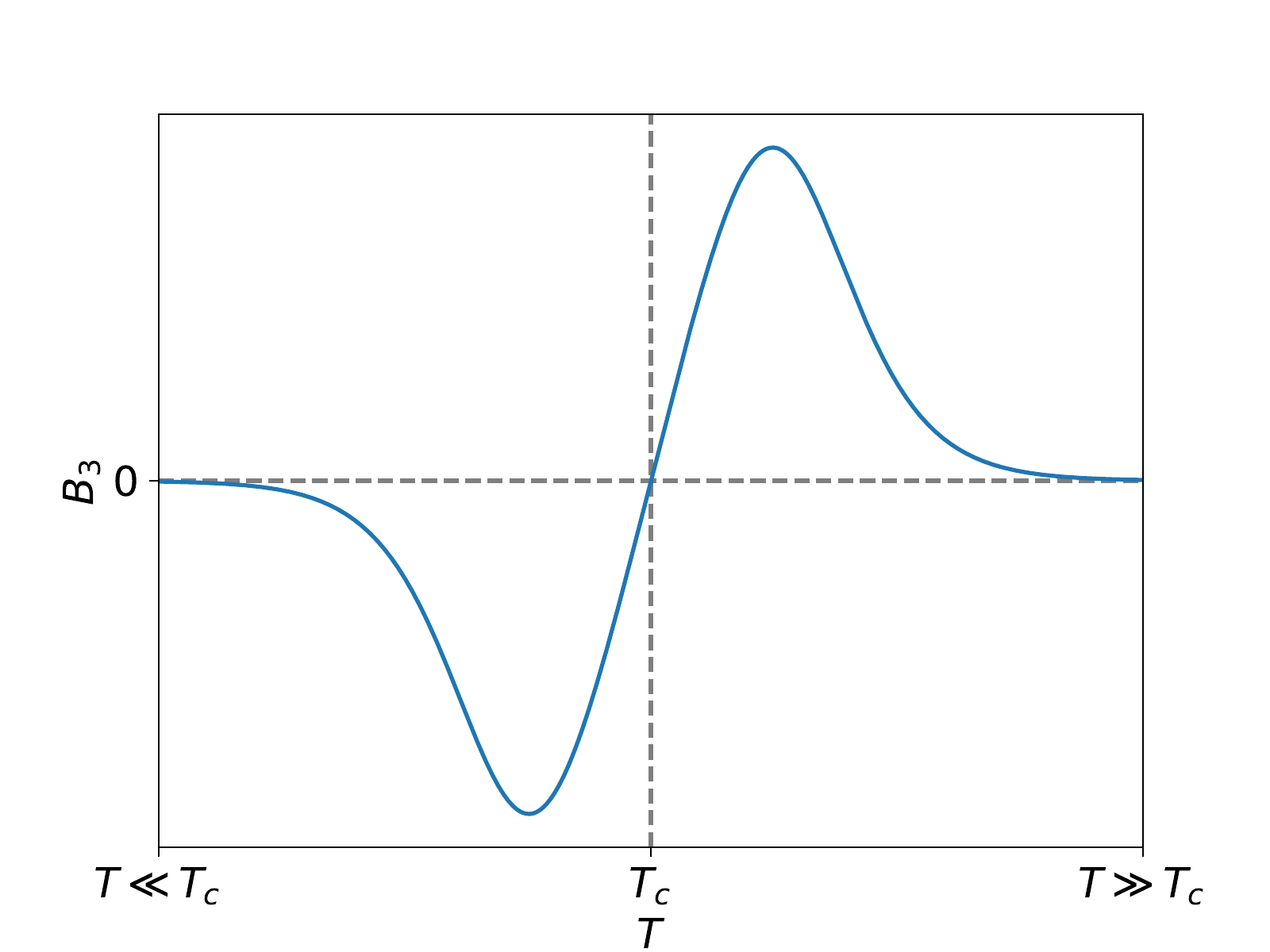}}
    \subfloat[Kurtosis\label{fig:moments_kurtosis}]{\includegraphics[width=0.5\textwidth]{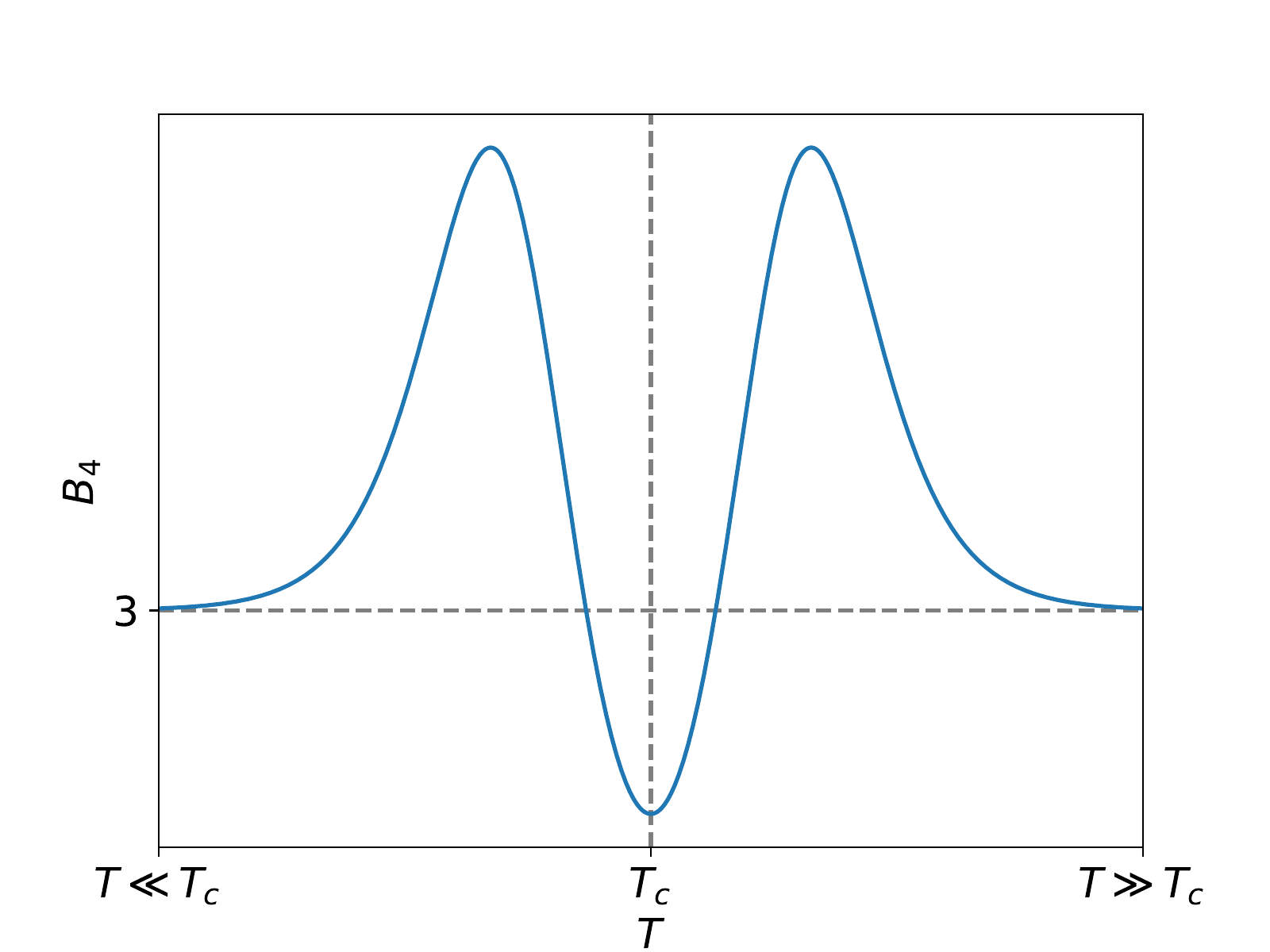}}
    \caption{Schematic behavior of the skewness $B_3(T)$ (\ref{fig:moments_skewness}) and kurtosis $B_4(T)$ (\ref{fig:moments_kurtosis}) of the order parameter as a function of temperature, modelled as in Ref.~\cite{cuteriQCDChiralPhase2018}.} 
    \label{fig:moments}
\end{figure}

The disconnected susceptibility is known to exhibit a peak at the critical temperature of the phase transition. As for the skewness  $B_3$ of the order parameter, its behavior is shown in Fig.~\ref{fig:moments_skewness}. It vanishes far away from the transition, then becomes non-zero, then vanishes at the critical temperature. Therefore, it can be used to locate the phase transition. The corresponding behavior of the kurtosis $B_4$ as a function of temperature is shown in Fig.~\ref{fig:moments_kurtosis}.

In order to extract information on the order of the phase transition, one can perform a finite-size scaling analysis of the kurtosis $B_4 (T_c)$ of the order parameter at the critical temperature. The value of the kurtosis on the transition in the thermodynamical limit is given by
\begin{equation}
    \lim_{N_\sigma \to \infty} B_4(T_c, m, N_\sigma) = 
    \begin{cases}
        1 & 1^{st} \text{ order} \\
        1.604 & 2^{nd} \text{ order } Z_2 \\
        3 & \text{crossover}
    \end{cases}
\end{equation}
and it can be used to distinguish the first-order case, the crossover case and the second-order $Z_2$ line, which is in the universality class of the three-dimensional Ising model.

\section{Lattice setup}

We start with the $N_f=3$ ensembles at $m_u = m_d = m_s$, and a pion mass of $m_\pi$ = 412 MeV produced by the Open Lattice initiative, varying the temperature by changing $N_\tau$. For ensemble generation the tree-level Symanzik improved gauge and the exponentiated Wilson Clover fermion actions are employed, together with the SMD update algorithm and the uniform norm stopping criterion for the force solvers. We employ twisted mass Hasenbusch preconditioning for $u$ and $d$ and the rational approximation for the $s$ quark. Reweighting is then needed to correct for the error due to the rational approximation, but we do not use a non-zero outer twisted mass parameter, which demands a twisted mass reweighting. The deflated Schwarz-preconditioned GCR solver \cite{luscherLocalCoherenceDeflation2007} is used whenever allowed by the choice of $N_\tau$, otherwise a conjugate gradient solver is used.

We consider lattices with spatial length such that $m_\pi L_\sigma \gtrsim 4$ in order to have relatively small finite size effects at the various lattice spacings.

\section{Preliminary results}

\begin{table}[h]
    \centering
    \begin{tabular}{c c c c c c}
        a [fm]   & $\kappa$  & $\beta$ & T [MeV] & $N_\tau$ & $N_\sigma$ \\ \hline \hline 
        0.12    & 0.1394305 & 3.685 & 274 & 6 & 16 \\
        &      & &  205 & 8 & 16 \\
        &      & &  164 & 10 & 16 \\ \hline
        0.094 & 0.138963 & 3.800 & 262 & 8 & 24 \\
        &      & & 210 &  10 & 24 \\
        &      & & 175 & 12 & 24 
    \end{tabular}
    \caption{Setups currently simulated and shown in the preliminary results.}
    \label{tab:runs}
\end{table}

\begin{figure}
    \centering
    \subfloat[$a = 0.12$ fm\label{fig:results_12}]{\includegraphics[width=0.5\textwidth]{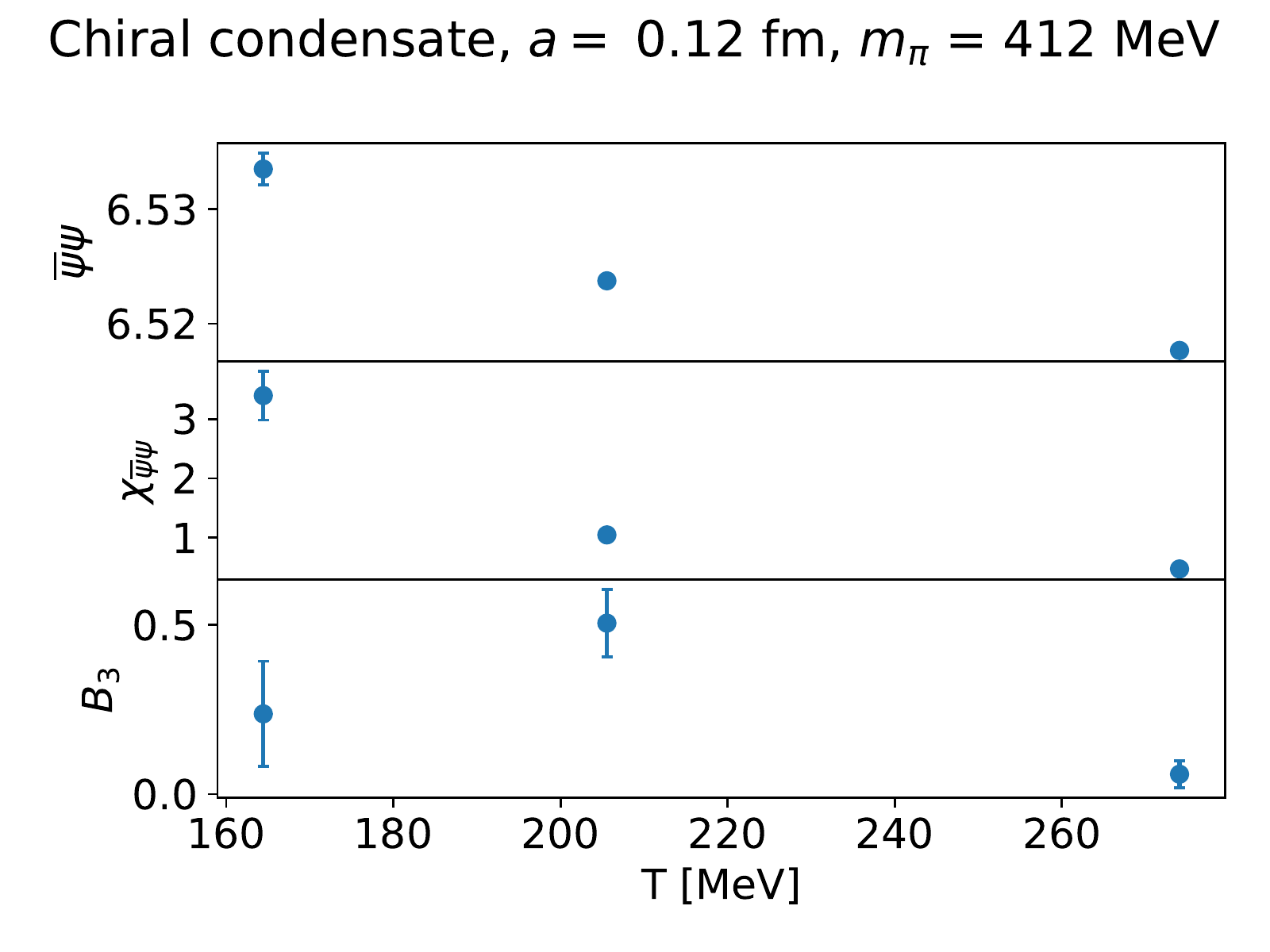}}
    \subfloat[$a = 0.094$ fm\label{fig:results_094}]{\includegraphics[width=0.5\textwidth]{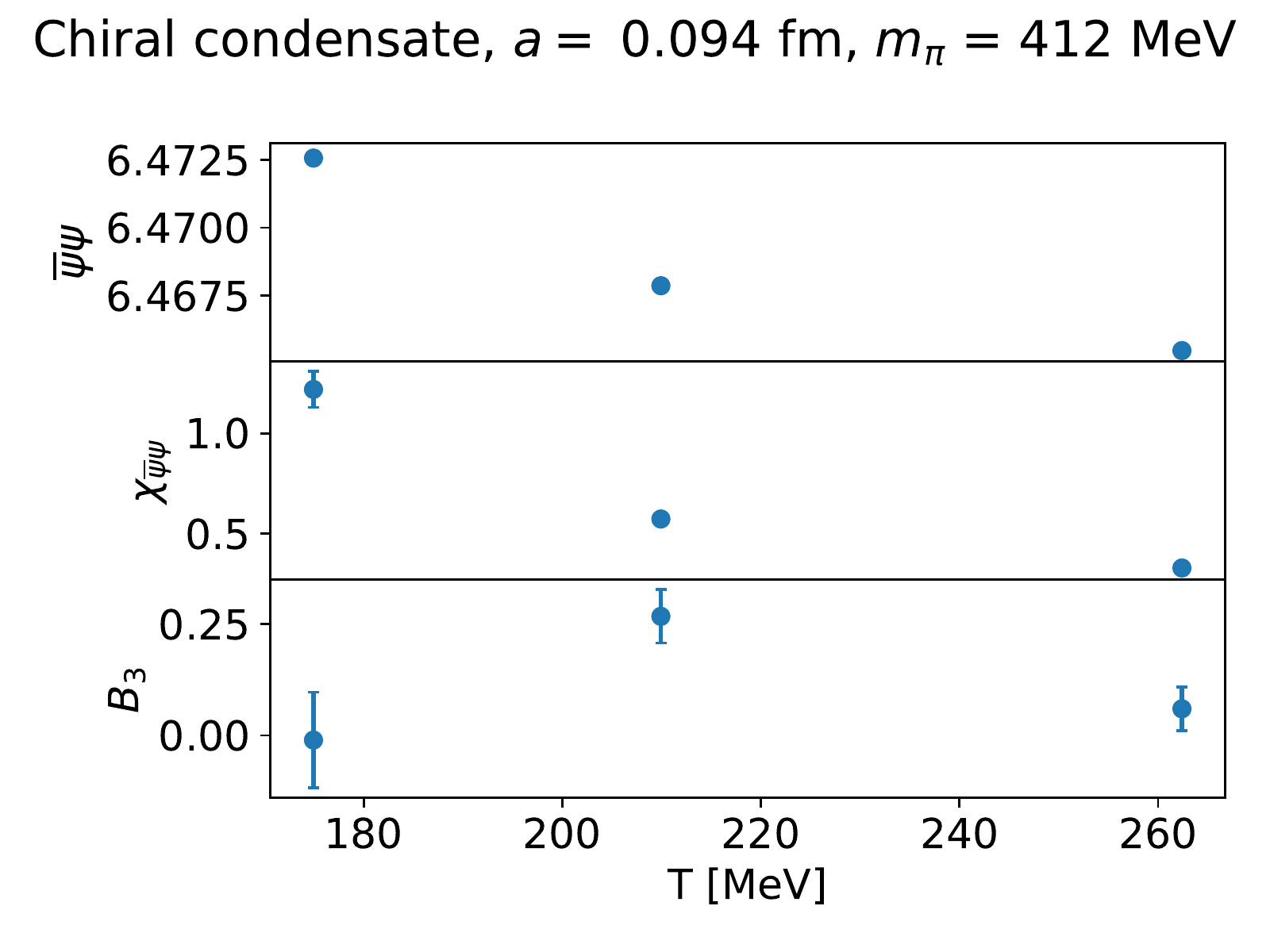}}
    \caption{Preliminary results of the light chiral condensate, its susceptibility and its skewness on the $a=0.12$ fm (\ref{fig:results_12}) and $a=0.094$ fm (\ref{fig:results_094}) lattices. Results are given in lattice units.}
    \label{fig:results}
\end{figure}

The currently simulated setups on the two coarsest lattices, corresponding to a lattice spacing of $0.12$ fm and $0.094$ fm, are shown in Tab.~\ref{tab:runs}. We show in Fig.~\ref{fig:results} the preliminary results (in lattice units) of the bare light chiral condensate, its susceptibility and its skewness $B_3$ on these lattices. We temporarily neglected the contribution of the reweighting factors for the rational approximation of the third quark flavour, as it is much smaller than the statistical errors. We notice that the skewness $B_3$ is positive, which hints that the critical temperature is smaller than the ones simulated. From the $a=0.12$ fm runs, we can estimate $T_c \lesssim 164$ MeV, while from the $a=0.094$ fm runs we notice that the skewness is compatible with zero already at $T_c \simeq 175$ MeV. 
Larger amounts of statistics are necessary to reduce the error bars, and simulations at smaller temperatures are needed to confirm the trend and to check that the peak of the susceptibility is consistent with the behavior of the skewness.

Due to the limited resolution in temperature, and to the limited amount of statistics, we cannot yet draw any conclusions from this analysis on the order of the transition at this pion mass, which we expect to be well in the crossover region. For our finest lattice ($a = 0.055$ fm), we expect to reach a resolution on the order of $20$ MeV around the critical temperature.

\section{Summary}

We performed an exploratory study of $N_f=2+1$ thermodynamics with stabilized Wilson Fermions, using as starting points the zero-temperature ensembles provided by the Open Lattice initiative. We started on the flavour symmetric line, at $m_\pi = 412$ MeV, and with the coarsest lattice spacings of $a=0.12 $ fm and $a=0.094$ fm, varying the temperature by changing $N_\tau$, and looked for the thermal phase transition by studying the susceptibility and moments of the chiral condensate.

Runs at smaller temperatures and more statistics are still needed to obtain a determination of $T_c$ at this pion mass, and finer lattice spacings are necessary both for a continuum extrapolation and for obtaining a finer resolution in temperature. Parts of this work serve as preparation for a future more detailed study, in which we aim to locate the $Z_2$ endpoint by varying $\beta$ at fixed $N_\tau$.

\section{Acknowledgments}

FC and RFB acknowledge the support by the State of Hesse within the Research Cluster ELEMENTS (Project ID 500/10.006). BB, FC and GE also acknowledge support by the Deutsche Forschungsgemeinschaft (DFG, German Research Foundation) via TRR 211 – project number 315477589. The simulations have been done on the Goethe-HLR cluster at the University of Frankfurt and we thank the computing staff for their support. AF acknowledges support by the Ministry of Science and Technology Taiwan (MOST) under grant 111-2112-M-A49-018-MY2. 

\bibliographystyle{JHEP}
\bibliography{bibliography.bib}

\end{document}